\documentstyle[11pt,twoside,epsfig]{article}
\newcommand{\beq}{\begin{equation}}
\newcommand{\eeq}{\end{equation}}
\newcommand{\beqa}{\begin{eqnarray}}
\newcommand{\eeqa}{\end{eqnarray}}
\newcommand{\beqan}{\begin{eqnarray*}}
\newcommand{\eeqan}{\end{eqnarray*}}
\newcommand{\ba}{\begin{array}}
\newcommand{\ea}{\end{array}}
\newcommand{\no}{\nonumber}

\newcommand{\ol}{\overline}
\newcommand{\ra}{\rightarrow}

\newcommand{\ve}{\varepsilon}

\newcommand{\dg}{\dagger}

\newcommand{\wh}{\widehat}

\newcommand{\cL}{{\cal L}}
\newcommand{\M}{{\cal M}}

\newcommand{\st}{\stackrel}
\newcommand{\dfrac}{\displaystyle \frac}
\newcommand{\del}{\partial}

\newcommand{\Fsl}{\not\!\!}

\normalsize
\begin{document}
\begin{titlepage}
\begin{flushright}
UWThPh-1995-22\\
July 1995\\
hep-ph/9508204
\end{flushright}
\vspace{2.5cm}
\begin{center}
{\Large \bf Low--Energy Expansion of the \\[5pt]
Pion--Nucleon Lagrangian*}\\[40pt]
G. Ecker$^1$ and M. Moj\v zi\v s$^{1,2}$

\vspace{1cm}

${}^{1)}$ Institut f\"ur Theoretische Physik, Universit\"at Wien\\
Boltzmanngasse 5, A--1090 Wien, Austria \\[10pt]

${}^{2)}$ Department of Theoretical Physics, Comenius University\\
Mlynsk\'a dolina, SK--84215 Bratislava, Slovakia

\vfill
{\bf Abstract} \\
\end{center}
\noindent
The renormalized pion--nucleon Lagrangian is calculated to $O(p^3)$
in heavy baryon chiral perturbation theory. By suitably chosen
transformations of the nucleon field, the Lagrangian is brought to
a standard form.

\vfill
\noindent * Work supported in part by FWF (Austria), Project No. P09505--PHY,
by HCM, EEC--Contract No. CHRX--CT920026 (EURODA$\Phi$NE) and by
the Austrian-Slovak Exchange Agreement
\end{titlepage}

\paragraph{1.} The effective field theory of the pion--nucleon interaction
in the one--nucleon sector can be formulated in terms of a chiral
Lagrangian \cite{GSS88}
\beq
\cL_{\pi N} = \cL_{\pi N}^{(1)} + \cL_{\pi N}^{(2)} + \cL_{\pi N}^{(3)}
+ \ldots.
\label{LMB}
\eeq
The chiral counting rules for constructing the Lagrangians
$\cL_{\pi N}^{(n)}$ of $O(p^n)$ are different for pions and nucleons
because the nucleon mass does not vanish in the chiral limit. Unlike
in the meson sector, there is no direct correspondence between the
chiral expansion and the loop expansion with the Lagrangian (\ref{LMB}).

Heavy baryon chiral perturbation theory (HBCHPT) \cite{JM91} eliminates
this drawback at the price of introducing a special reference frame
characterized by a time--like unit four--vector $v$. In constructing the
chiral Lagrangian of HBCHPT, care must be exercised to satisfy Lorentz
invariance. The key observation in this context is that in the presence of
external objects like the four--vector $v$ Lorentz covariance
alone is not sufficient to guarantee Lorentz invariance.

The purpose of this letter is to construct the complete renormalized
chiral Lagrangian for the pion--nucleon system to $O(p^3)$ in HBCHPT.
To ensure Lorentz invariance, we start from the fully
relativistic Lagrangian (\ref{LMB}) to perform the frame--dependent
decomposition of HBCHPT. With such a procedure, reparametrization
invariance \cite{LM92} is automatically fulfilled.

Both the non--relativistic decomposition and the loop expansion generate
so--called equation--of--motion terms. By successive transformations of
the (frame--dependent) nucleon field, those terms will be eliminated
 from the chiral Lagrangian. After these transformations, the final
Lagrangian yields directly all one--particle--irreducible vertices in the
usual manner.

\paragraph{2.} Our starting point is QCD with two massless quarks $u$,
$d$ (all others are massive) coupled to external fields \cite{GL84}:
\beq
\cL = \cL_{\rm QCD}^0 + \bar q \gamma^\mu \left(v_\mu + \frac{1}{3}
v_\mu^{(s)} + \gamma_5 a_\mu \right) q - \bar q (s - i \gamma_5 p)q,
\qquad q = \left( \ba{c} u \\ d \ea \right).
\label{QCD}
\eeq
The isotriplet vector and axial--vector fields $v_\mu$, $a_\mu$ are
traceless hermitian and $s$, $p$ are hermitian matrix fields. The
isosinglet vector field $v_\mu^{(s)}$ is needed to generate the
electromagnetic current. Explicit chiral symmetry breaking is implemented
by setting $s = \M = \mbox{diag }(m_u,m_d)$.

The Lagrangian (\ref{QCD}) exhibits a local chiral symmetry
$SU(2)_L \times SU(2)_R \times U(1)_V$ that breaks spontaneously to
$SU(2)_V \times U(1)_V$. It is realized non--linearly on the Goldstone
pion fields $\phi$ and on the nucleon field $\Psi$ \cite{CCWZ69}:
\beqa
\label{trafo}
&& u(\phi) \st{g}{\ra} g_R u(\phi) h(g,\phi)^{-1} =
h(g,\phi) u(\phi) g_L^{-1} \no \\
&& \Psi = {p \choose n} \st{g}{\ra} h(g,\phi) \Psi \\
&& g = (g_L,g_R) \in SU(2)_L \times SU(2)_R~. \no
\eeqa
The matrix field $u(\phi)$ is an element of the chiral coset space and
the compensator field $h(g,\phi) \in SU(2)_V$ characterizes the
non--linear realization. The baryon number symmetry
$U(1)_V$ acts only on the nucleon field.

The following ingredients are needed to construct the chiral Lagrangian
(\ref{LMB}):
\beqa
\label{ing}
u_\mu &=& i \{ u^\dg(\partial_\mu - i r_\mu)u -
u(\partial_\mu - i \ell_\mu) u^\dg\} \no \\
\Gamma_\mu &=& \frac{1}{2} \{ u^\dg (\partial_\mu - i r_\mu)u +
u (\partial_\mu - i \ell_\mu) u^\dg \} \no \\
\chi_\pm &=&  u^\dg \chi u^\dg \pm u \chi^\dg u ~,\qquad
\chi= 2 B (s+ip)  \\
f_\pm^{\mu\nu} &=& u F_L^{\mu\nu} u^\dg \pm u^\dg F_R^{\mu\nu} u \no \\
v_{\mu\nu}^{(s)} &=& \partial_\mu v_\nu^{(s)} - \partial_\nu v_\mu^{(s)}.
\no
\eeqa
Here, $B$ is a parameter of the meson Lagrangian of $O(p^2)$
\cite{GL84} and $r_\mu = v_\mu + a_\mu$, $\ell_\mu = v_\mu - a_\mu$
are the external gauge fields with associated non--Abelian field strengths
\beqa
F_R^{\mu\nu} &=& \partial^\mu r^\nu - \partial^\nu r^\mu - i[r^\mu,r^\nu] \\
F_L^{\mu\nu} &=& \partial^\mu \ell^\nu - \partial^\nu \ell^\mu -
i [\ell^\mu,\ell^\nu]~. \no
\eeqa
With a covariant derivative
\beq
\nabla_\mu = \partial_\mu + \Gamma_\mu - i v_\mu^{(s)}, \label{cov}
\eeq
the lowest--order chiral Lagrangian in (\ref{LMB}) takes the form
\cite{GSS88}
\beq
\cL_{\pi N}^{(1)} = \bar \Psi \left(i \not\!\nabla - m + \frac{g_A}{2}
\not\!u \gamma_5\right)\Psi, \label{LMB1}
\eeq
where $m$, $g_A$ are the nucleon mass and the axial--vector coupling
constant in the chiral limit.

HBCHPT rewrites the Lagrangian (\ref{LMB}) in terms of velocity--dependent
fields $N_v$, $H_v$  defined as \cite{Ge90}
\beqa
\label{vdf}
N_v(x) &=& \exp[i m v \cdot x] P_v^+ \Psi(x)  \\
H_v(x) &=& \exp[i m v \cdot x] P_v^- \Psi(x) \no \\
P_v^\pm &=& \frac{1}{2} (1 \pm \not\!v)~, \qquad v^2 = 1 ~. \no
\eeqa
The pion--nucleon Lagrangian takes the form
\beq
\cL_{\pi N} = \bar N_v A N_v + \bar H_v B N_v + \bar N_v \gamma^0 B^\dg
\gamma^0 H_v - \bar H_v C H_v \label{piN}
\eeq
where $A$, $B$, $C$ are mesonic field operators with a straightforward
chiral expansion, e.g.,
\beqa
\label{A1}
A &=& A_{(1)} + A_{(2)} + A_{(3)} + \ldots \no \\
A_{(1)} &=& iv \cdot \nabla + g_A S \cdot u
\eeqa
with the spin matrix $S^\mu = i \gamma_5 \sigma^{\mu\nu} v_\nu/2$.
In the generating functional of Green functions \cite{GSS88}, the fields
$H_v$ are integrated out to produce a non--local action in terms of the
fields $N_v$ \cite{MRR92,BKKM92,Ec94a}
\beq
S_{\pi N} = \int d^4x \; \wh\cL_{\pi N} = \int d^4x \; \bar N_v
(A + \gamma^0 B^\dg \gamma^0 C^{-1} B)N_v~.
\eeq
Expanding $C^{-1}$ in inverse powers of the nucleon mass, one arrives
finally at the chiral Lagrangian of HBCHPT. Up to $O(p^3)$, its general
form is given by
\beqa
\label{ABC}
\lefteqn{A + \gamma^0 B^\dg \gamma^0 C^{-1} B = A_{(1)} } \no \\
&& \mbox{} + A_{(2)} + \frac{1}{2m} \gamma^0 B^\dg_{(1)} \gamma^0
B_{(1)} \no \\
&& \mbox{} + A_{(3)} + \frac{1}{2m} (\gamma^0 B^\dg_{(2)} \gamma^0
B_{(1)} + \gamma^0 B^\dg_{(1)} \gamma^0 B_{(2)}) - \frac{1}{4m^2}
\gamma^0 B^\dg_{(1)} \gamma^0 (iv \cdot \nabla + g_A S \cdot u)
B_{(1)} \no \\
&& \mbox{} + O(p^4).
\eeqa
The chiral Lagrangian of $O(p)$ is determined by the operator
$A_{(1)}$ in (\ref{A1}):
\beq
\wh \cL_{\pi N}^{(1)} = \bar N_v(iv \cdot \nabla + g_A S \cdot u) N_v~.
\label{piN1}
\eeq
In order to obtain the Lagrangians of $O(p^2)$ and $O(p^3)$, we are
going to construct the most general operators $A_{(2)}$, $A_{(3)}$
and $B_{(2)}$ compatible with chiral symmetry, parity, charge conjugation
and Lorentz invariance.

\paragraph{3.} The relativistic Lagrangian of $O(p^2)$ was considered
in Refs. \cite{GSS88,Kr90}. Our procedure is slightly different since
we include all field monomials of the form $\bar \Psi P_2 \Psi$
that contribute to $A_{(2)}$ and/or $B_{(2)}$. The complete list of
independent operators $P_2$ is as follows:
\beqa
\label{P2}
\langle u_\mu u^\mu\rangle, \; \langle u_\mu u_\nu\rangle \nabla^\mu
\nabla^\nu + {\rm h.c.}, \; \langle \chi_+\rangle, \; \chi_+ - \frac{1}{2}
\langle \chi_+ \rangle &\ra& A_{(2)} \no \\
i \sigma^{\mu\nu} u_\mu u_\nu, \; \sigma^{\mu\nu} f_{+\mu\nu}, \;
\sigma^{\mu\nu} v_{\mu\nu}^{(s)}, \; \langle u_\mu u_\nu \rangle
i \gamma^\mu \nabla^\nu + {\rm h.c.} &\ra& A_{(2)}, B_{(2)} \\
\chi_-\gamma_5, \; \langle \chi_- \rangle \gamma_5, \;
i \gamma_5 [\nabla_\mu,u_\nu] \{\nabla^\mu,\nabla^\nu\} + {\rm h.c.} &\ra&
B_{(2)} \no
\eeqa
where $\langle \ldots\rangle$ stands for the trace. We refer to the paper
of Krause \cite{Kr90} for a thorough exposition of how to arrive at a
minimal set of such operators. Note in particular that all
equation--of--motion terms proportional to $(i\Fsl\nabla - m)\Psi$
can be eliminated by a transformation of the nucleon field $\Psi$.

The two pieces of $O(p^2)$ in (\ref{ABC}) are then found \cite{BKKM92}
to be
\beqa
\label{A2}
A_{(2)} &=& \frac{c_1}{m} \langle \chi_+\rangle + \frac{c_2}{m}
(v \cdot u)^2 + \frac{c_3}{m} u \cdot u + \frac{c_5}{m}
\left( \chi_+ - \frac{1}{2} \langle \chi_+\rangle \right) \no \\
&& \mbox{} + \frac{1}{m} \ve^{\mu\nu\rho\sigma} v_\rho S_\sigma
[i c_4 u_\mu u_\nu + c_6 f_{+\mu\nu} + c_7 v^{(s)}_{\mu\nu}]
\eeqa
and
\beqa
\label{B1B1}
\lefteqn{\frac{1}{2m} \gamma^0 B^\dg_{(1)} \gamma^0 B_{(1)} = \frac{1}{2m}
\{ (v \cdot \nabla)^2 - \nabla \cdot \nabla - ig_A \{S \cdot \nabla,
v \cdot u\} } \no \\
&& \mbox{} - \frac{g_A^2}{4} (v \cdot u)^2 + \frac{1}{2}
\ve^{\mu\nu\rho\sigma} v_\rho S_\sigma [i u_\mu u_\nu + f_{+\mu\nu}
+ 2 v^{(s)}_{\mu\nu}]\}~.
\eeqa
The present status of the scale--independent, dimensionless low--energy
constants (LECs) $c_1, \ldots, c_7$ is summarized in Refs. \cite{Ec95,BKM95}.

In order to incorporate all 1PI vertices in the Lagrangian, the operator
$(v \cdot \nabla)^2/2m$ in (\ref{B1B1}) can be eliminated by a field
transformation \footnote{The heavy--nucleon propagator
in the presence of external fields is given by the inverse of
$v \cdot \nabla$.}. Up to
$O(p^3)$, we will encounter such equation--of--motion terms of the
following variety:
\beq
\cL_{\rm EOM} = \bar N_v \{ X(iv \cdot \nabla)^3 + iv \cdot \st{\gets}{\nabla}
Y iv \cdot \nabla + Z iv \cdot \nabla - iv \cdot \st{\gets}{\nabla} Z^\dg\}
N_v~. \label{EOM}
\eeq
The mesonic operators $Y = Y^\dg$ and $Z$ are at most of $O(p)$ and $O(p^2)$,
respectively and $X = X^*$ is a constant. Applying a nucleon field
transformation
\beqa
N_v &=& \left\{ 1 - \frac{X}{2} (iv \cdot \nabla)^2 + \frac{1}{2}
(Y + g_A X S \cdot u)iv \cdot \nabla + \frac{g_A}{2} X[iv \cdot \nabla,
S \cdot u] \right. \no \\
&& \left.\mbox{} - \frac{g^2_A}{2} X(S \cdot u)^2 - \frac{g_A}{2} Y S \cdot u
- Z^\dg \right\} N'_v
\label{nft}
\eeqa
to $\wh \cL_{\pi N}^{(1)}$ in (\ref{piN1}) and $\cL_{\rm EOM}$ in
(\ref{EOM}) eliminates the equation--of--motion terms and generates
the following Lagrangian of the same chiral order as $\cL_{\rm EOM}$:
\beqa
\cL_{\rm induced} &=& \ol{N'_v} \{
- g^3_A X(S \cdot u)^3 + \frac{g^2_A}{2} X[S \cdot u,[iv \cdot \nabla,
S \cdot u]]  \no \\
&& \left.\mbox{} - g^2_A S \cdot u Y S \cdot u - g_A(ZS\cdot u +
S \cdot u Z^\dg) \right\} N'_v~.
\label{Lind}
\eeqa
Of course, the transformation (\ref{nft}) must be applied to the complete
Lagrangian to the order considered and it will therefore generate
additional terms. For the case at hand (with $X=Z=0$, $Y=1/2m$), the
complete Lagrangian of $O(p^2)$ in HBCHPT assumes its final form
\beqa
\label{piN2}
\wh \cL_{\pi N}^{(2)} &=& \bar N_v\left( - \frac{1}{2m} (\nabla \cdot
\nabla + ig_A \{S \cdot \nabla, v \cdot u\}) \right. \no \\
&& \mbox{} + \frac{a_1}{m} \langle u \cdot u\rangle +
\frac{a_2}{m} \langle (v \cdot u)^2\rangle +
\frac{a_3}{m} \langle \chi_+\rangle +
\frac{a_4}{m} \left( \chi_+ - \frac{1}{2} \langle \chi_+\rangle \right)
\no \\
&& \left. \mbox{} + \frac{1}{m} \ve^{\mu\nu\rho\sigma} v_\rho S_\sigma
[i a_5 u_\mu u_\nu + a_6 f_{+\mu\nu} + a_7 v_{\mu\nu}^{(s)}]\right) N_v~.
\eeqa
To facilitate comparison, we display the relations between the LECs $a_i$ and
the previously defined $c_i$:
\beq
\label{ai}
\ba{llll}
a_1 = \dfrac{c_3}{2} + \dfrac{g^2_A}{16}, \qquad &
a_2 = \dfrac{c_2}{2} - \dfrac{g^2_A}{8}, \qquad &
a_3 = c_1, \qquad & a_4 = c_5, \\[10pt]
a_5 = c_4 + \dfrac{1-g^2_A}{4},\qquad & a_6 = c_6 + \dfrac{1}{4}, &
a_7 = c_7 + \dfrac{1}{2}~. \ea
\eeq
The LECs $a_6$, $a_7$ are related to the nucleon magnetic moments in the
chiral limit:
\beqa
\label{nmm}
a_6 &=& \frac{\mu_v}{4} = \frac{1}{4} (\mu_p - \mu_n) \no \\
a_7 &=& \frac{\mu_s}{2} = \frac{1}{2} (\mu_p + \mu_n)~.
\eeqa

The two terms in (\ref{piN2}) with fixed coefficients are a manifestation
of the difference between Lorentz invariance and covariance and they are
of course consistent with reparametrization invariance \cite{LM92}.
In particular, they generate the Thomson limit in nucleon Compton scattering
\cite{BKKM92} and the $O(p^2)$ amplitude for pion photoproduction at
threshold \cite{BGKM91}.

The field transformation (\ref{nft}) produces also terms of $O(p^3)$
and even higher orders. The terms of $O(p^3)$ will be included in
the final Lagrangian (\ref{piN3}).

\paragraph{4.} As shown in Eq. (\ref{ABC}), there are three different
pieces in the chiral Lagrangian $\wh \cL_{\pi N}^{(3)}$ of $O(p^3)$.
The term $A_{(3)}$ is obtained from the non--relativistic decomposition
of the relativistic Lagrangian $\cL_{\pi N}^{(3)}$. We refrain from writing
down the Lagrangian $\cL_{\pi N}^{(3)}$ as it coincides with the $SU(2)$
version of Krause's Lagrangian of $O(p^3)$ \cite{Kr90}. Some of
the couplings in $A_{(3)}$ must absorb the divergences of the one--loop
functional with vertices from $\wh \cL_{\pi N}^{(1)}$ and from the
lowest--order mesonic Lagrangian. Those divergences were calculated in
Ref. \cite{Ec94b}. Since they contain seven independent equation--of--motion
terms of the type (\ref{EOM}), we perform another field transformation to get
rid of those terms. At $O(p^3)$, this transformation modifies the coefficients
of the remaining monomials. In addition, this transformation also influences
the divergence structure at higher orders. In particular, applying it to
$\wh \cL_{\pi N}^{(2)}$ in (\ref{piN2}) generates divergent terms of
$O(p^4)$ that have to be included in the full (one--loop) functional
of $O(p^4)$ to be calculated eventually.

Here, we concentrate on the Lagrangian of $O(p^3)$. Altogether, we find 22
independent terms for $A_{(3)}$. However, two more terms with arbitrary
coupling constants emerge from the terms in $\wh \cL_{\pi N}^{(3)}$
involving $B_{(2)}$ [cf. Eq. (\ref{P2})]. The complete list of
independent field monomials of $O(p^3)$ is displayed in Table~1 (the
additional terms due to $B_{(2)}$ correspond to $i = 19, 20$).

\renewcommand{\arraystretch}{1.1}
\begin{table}
\caption{Field monomials of $O(p^3)$ in the Lagrangian
(\protect\ref{piN3}) and their $\beta$--functions.}
$$
\begin{tabular}{|r|c|c|} \hline
i  & $O_i$ & $\beta_i$ \\ \hline
1  & $i[u_\mu,[v \cdot \nabla, u^\mu]]$ & $-g^4_A/6$ \\
2  & $i[u_\mu,[\nabla^\mu, v \cdot u]]$ & $- (1 + 5 g^2_A)/12$ \\
3  & $i[v \cdot u, [v \cdot \nabla, v \cdot u]]$ & $(3 + g^4_A)/6$ \\
4  & $i \langle u_\mu v \cdot u\rangle \nabla^\mu + $ h.c. & $0$ \\
5  & $i v_\lambda \ve^{\lambda \mu \nu \rho} \langle u_\mu u_\nu u_\rho
\rangle$ & $0$ \\
6 & $[\chi_-, v \cdot u]$ & $(1+5g^2_A)/24$ \\
7  & $[\nabla^\mu,f_{+\mu\nu}]v^\nu$ & $-(1+5g^2_A/6$ \\
8  & $\partial^\mu v_{\mu\nu}^{(s)} v^\nu$ & $0$ \\
9  & $\ve^{\mu\nu\rho\sigma} \langle f_{+\mu\nu} u_\rho \rangle
v_\sigma$ & $0$ \\
10 & $\ve^{\mu\nu\rho\sigma} v_{\mu\nu}^{(s)} u_\rho v_\sigma$ & $0$ \\
11 & $S \cdot u \langle u \cdot u \rangle$ & $g_A(1+5g^2_A + 4g^4_A)/2$ \\
12 & $u_\mu S_\nu \langle u^\mu u^\nu\rangle$
& $ g_A (3 - 9 g^2_A + 4 g^4_A)/6$ \\
13 & $S \cdot u \langle (v \cdot u)^2\rangle$ & $ - g_A(2 + g^2_A +
2 g^4_A)$ \\
14 & $v \cdot u S_\mu \langle u^\mu v \cdot u\rangle$
& $g^3_A + 2g^5_A/3 $ \\
15 & $\ve^{\mu\nu\rho\sigma} v_\rho S_\sigma \langle [v \cdot \nabla,
u_\mu]u_\nu \rangle$ & $g^4_A/3$ \\
16 & $\ve^{\mu\nu\rho\sigma} v_\rho S_\sigma \langle u_\mu [\nabla_\nu,
v \cdot u]\rangle$ & $0$ \\
17 & $S \cdot u \langle \chi_+\rangle$ & $g_A/2 + g^3_A$ \\
18 & $S^\mu \langle u_\mu \chi_+ \rangle $ & $0$ \\
19 & $i S^\mu [\nabla_\mu,\chi_-]$  & $0$ \\
20 & $i S^\mu \langle \partial_\mu \chi_-\rangle $ & $0$ \\
21 & $i S^\mu v^\nu [f_{+\mu\nu}, v \cdot u]$ & $g_A + g^3_A $ \\
22 & $i S^\mu [f_{+\mu\nu},u^\nu]$  & $- g^3_A$ \\
23 & $S^\mu [\nabla^\nu,f_{-\mu\nu}]$ & 0 \\
24 & $\ve^{\mu\nu\rho\sigma} S_\mu \langle u_\nu f_{-\rho\sigma}\rangle$
& 0 \\ \hline
\end{tabular}
$$
\end{table}

Calculation of the remaining terms in (\ref{ABC}) of $O(p^3)$ with
coefficients defined at $O(p)$ or $O(p^2)$ is straightforward. Once again,
one encounters equation--of--motion terms involving $v \cdot \nabla N_v$.
Another basis transformation (not affecting the Lagrangian to $O(p^2)$,
of course) brings the Lagrangian of $O(p^3)$ into its final form
\beqa
\label{piN3}
\wh \cL_{\pi N}^{(3)} &=& \bar N_v \left( \frac{g_A}{8 m^2}
[ \nabla_\mu,[\nabla^\mu,S \cdot u]] + \frac{1}{2m^2} \left[
\left\{ i \left( a_5 - \frac{1-3g^2_A}{8}\right) u_\mu u_\nu
 \right. \right. \right. \no \\
&& \mbox{} + \left. \left( a_6 - \frac{1}{8}\right) f_{+\mu\nu} +
\left( a_7 - \frac{1}{4}\right) v_{\mu\nu}^{(s)} \right\}
\ve^{\mu\nu\rho\sigma} S_\sigma i \nabla_\rho + \frac{g_A}{2}
S \cdot \nabla u \cdot \nabla \no \\
&& \mbox{} - \left. \frac{g^2_A}{8} \{v \cdot u,u_\mu\}
\ve^{\mu\nu\rho\sigma} v_\rho S_\sigma \nabla_\nu - \frac{ig_A}{16}
\ve^{\mu\nu\rho\sigma} f_{-\mu\nu} v_\rho \nabla_\sigma + {\rm h.c.}\right]
\no \\
&& \mbox{} + \left.\frac{1}{(4\pi F)^2} \sum_{i=1}^{24} b_i O_i \right)
N_v~.
\eeqa
Some of the $b_i$ are divergent to absorb the divergences of the one--loop
functional of $O(p^3)$ \cite{Ec94b}. Decomposing them in the standard way
(using dimensional regularization) as
\beqa
b_i &=& b_i^r(\mu) + (4\pi)^2 \beta_i \Lambda(\mu) \\
\Lambda(\mu) &=& \frac{\mu^{d-4}}{(4\pi)^2} \left\{ \frac{1}{d-4} -
\frac{1}{2} [\ln 4\pi + 1 + \Gamma'(1)] \right\} \no
\eeqa
gives rise to 24 measurable renormalized LECs $b_i^r(\mu)$. The
$\beta_i$ are dimensionless functions of $g_A$ listed in Table~1 and
they govern the scale dependence of the $b_i^r(\mu)$:
\beq
b_i^r(\mu) = b_i^r(\mu_0) + \beta_i \ln \frac{\mu_0}{\mu}~.
\eeq

It is important to realize that the LECs $b_i^r(\mu)$ and their
$\beta$--functions depend on the choice of the mesonic Lagrangian
of $O(p^4)$. In addition to the action for the Lagrangian (\ref{piN3}),
the generating functional of $O(p^3)$ consists of four pieces corresponding
to two irreducible and two reducible diagrams \cite{GSS88,Ec94b}. It
can be shown by explicit calculation that the sum of the two reducible
diagrams (reproduced in Fig. 1) is finite and scale independent
if the mesonic Lagrangian $\cL_{4}^{\rm GSS}$ of
Ref.~\cite{GSS88} is used for the functional tree diagram in Fig. 1.
The divergences of the one--loop functional are therefore all contained
in the irreducible diagrams \cite{Ec94b} and they give rise to
the $\beta_i$ of Table 1.

\begin{figure}
\centerline{\epsfig{file=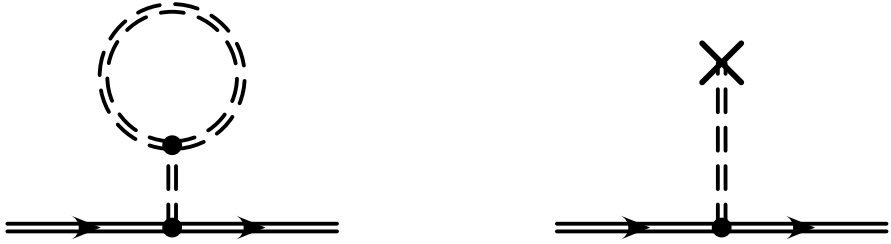,height=3cm}}
\caption{Reducible functional diagrams of $O(p^3)$. The full (dashed)
lines denote the nucleon (meson) propagators. The propagators and the
vertices have the full tree--level structure attached as functionals
of the external fields. The cross denotes a vertex of $O(p^4)$
 from the mesonic Lagrangian $\cL_4$.}
\end{figure}

However, the mesonic Lagrangian of $O(p^4)$ is not unique. One
can always use the equations of motion to transform it into different,
but physically equivalent forms. In fact, the original Lagrangian
$\cL_{4}^{\rm GL}$ \cite{GL84} differs from $\cL_{4}^{\rm GSS}$ by
precisely such an equation--of--motion term:
\beq
\cL_{4}^{\rm GL} - \cL_{4}^{\rm GSS} = \dfrac{l_4}{16} \left(
4 i\langle u_\mu \chi_-^\mu\rangle - 4 \langle \chi_+ u_\mu u^\mu\rangle
-2 \langle \chi_-^2\rangle + \langle \chi_-\rangle^2 \right)
\label{Ldiff}
\eeq
$$
\chi_-^\mu = u^\dg D^\mu \chi u^\dg - u D^\mu \chi^\dg u~,\qquad
D^\mu \chi = \del^\mu \chi -i r^\mu \chi + i \chi \ell^\mu~.
$$
Applying partial integration and the equations of motion
\beq
[\nabla^\mu, u_\mu] = \dfrac{i}{2}\chi_- -\dfrac{i}{4}\langle\chi_-\rangle~,
\label{MEOM}
\eeq
the difference of the corresponding actions can be shown to vanish.
The problem, which also appears in the calculation of the
mesonic functional of $O(p^6)$ \cite{Ec95}, is that use of the equations
of motion (\ref{MEOM}) in the construction of $\cL_{4}$ does not
commute with calculating vertices from $\cL_{4}$. To obtain the relevant
vertex in the tree diagram of Fig. 1, one expands
the action to first order in the fluctuation variables $\xi$ around
the classical solution. In the notation of Ref.~\cite{Ec94b},
one finds
\beq
S_{4}^{\rm GL} - S_{4}^{\rm GSS} = \dfrac{i l_4}{4} (\chi_- -
\dfrac{1}{2}\langle \chi_-\rangle)_j (d_\mu d^\mu + \sigma)_{ji}\xi_i
+O(\xi^2)~,
\eeq
where all quantities except $\xi$ are to be taken at the
classical solution. The meson propagator in the tree diagram of Fig. 1
is given by the inverse of the differential operator
$d_\mu d^\mu + \sigma$ (times a normalization factor $2/F^2$).
Thus, as expected on general grounds \cite{Ec95}, the difference
between the two functionals is a local action. Recalling the functional
nucleon--nucleon--meson vertex \cite{Ec94b} appearing in the
diagram in question,
\beq
V_i = \dfrac{i}{4\sqrt{2}} [v\cdot u,\tau_i] - \dfrac{g_A}{\sqrt{2}}
\tau_j S^\mu d_{\mu,ji}~,
\eeq
the difference between the tree--diagram functionals for the two
choices of $\cL_{4}$ is
\beq
Z_{\rm tree}^{\rm GL} - Z_{\rm tree}^{\rm GSS} = \dfrac{l_4}{8 F^2}
\int d^4x \bar N_v \left( [\chi_-,v\cdot u] -4 i g_A S_\mu [\nabla^\mu,
\chi_-] + 2 i g_A S_\mu\langle \del^\mu \chi_-\rangle\right)N_v~.
\eeq
Since $l_4$ is divergent \cite{GL84}, the sum of the two
functional diagrams in Fig. 1 diverges for $\cL_{4}^{\rm GL}$ because
it is finite for $\cL_{4}^{\rm GSS}$. Therefore, the entries in
Table 1 must be modified if the mesonic Lagrangian $\cL_{4}^{\rm GL}$
is used:
\beq
\beta_6^{\rm GL}= - \dfrac{5(1-g_A^2)}{24}~, \qquad
\beta_{19}^{\rm GL} = g_A~,\qquad \beta_{20}^{\rm GL} = -
\dfrac{g_A}{2}~,
\label{bdiff}
\eeq
all other coefficients being identical to the ones in Table 1.

The main lesson from this analysis is that the LECs $b_i$ depend
in general on the choice of pion fields. For the two particular choices
$\cL_{4}^{\rm GSS}$ and $\cL_{4}^{\rm GL}$, one finds in addition
to (\ref{bdiff})
\beqa
b_6^{{\rm GL},r}(\mu) &=& b_6^{{\rm GSS},r}(\mu) - \dfrac{(4\pi)^2}{8}
l_4^r(\mu)\no\\
b_{19}^{{\rm GL},r}(\mu) &=& b_{19}^{\rm GSS} + \dfrac{(4\pi)^2}{2}
g_A l_4^r(\mu) \\
b_{20}^{{\rm GL},r}(\mu) &=& b_{20}^{\rm GSS} - \dfrac{(4\pi)^2}{4}
g_A l_4^r(\mu) \no
\eeqa
for the renormalized LECs. All other coupling constants remain unchanged.

\paragraph{5.} Although different parts of the Lagrangian (\ref{piN3})
have been used before, the complete Lagrangian of $O(p^3)$ is given here
for the first time. Our knowledge of the $b_i$ is still rather
limited. For an up--to--date account, we refer to Ref. \cite{BKM95}.

Of special interest are the terms in the Lagrangian (\ref{piN3}) with
coefficients defined at $O(p)$ or $O(p^2)$. As an example, consider
the terms with coefficients $a_6$, $a_7$ that contribute to the
spin--dependent amplitude in nucleon Compton scattering. Projecting out
the electromagnetic field produces an amplitude proportional to
\beq
k_3 = \frac{1}{2} (1+ \tau_3) \left[ \left(a_6 - \frac{1}{8}\right)
\tau_3 + \frac{1}{2} \left( a_7 - \frac{1}{4}\right) \right] =
\frac{1}{4} \left( \ba{cc} 1 + 2\kappa_p & 0 \\ 0 & 0 \ea \right)~,
\eeq
with $\kappa_p$ the anomalous magnetic moment of the proton (in the
chiral limit). There is an additional tree--level contribution
\cite{BKKM92} with two
vertices of $O(p^2)$, each proportional to [cf. Eq. (\ref{piN2})]
\beq
k_2 = a_6 \tau_3 + \frac{a_7}{2} = \frac{1}{2}
\left( \ba{cc} 1 + \kappa_p & 0 \\ 0 & \kappa_n \ea \right) .
\eeq
The leading contribution to the spin--dependent Compton amplitude
in the forward direction for
small photon energies is of $O(p^3)$ and it is proportional to
\beq
k^2_2 - k_3 = \frac{1}{4} \left( \ba{cc} \kappa^2_p & 0 \\ 0 & \kappa^2_n
\ea \right)
\eeq
in accordance with a classic low--energy theorem \cite{LGMG54}.

The spin--dependent nucleon Compton scattering amplitude
is an example for the class of amplitudes in the one--nucleon
sector that are insensitive to the LECs $b_i$. All such amplitudes
are uniquely determined by LECs of at most $O(p^2)$ with possible
loop contributions being necessarily finite. Another prominent example is
the electric dipole amplitude $E_{0+}$ for $\pi^0$ photoproduction at
threshold that receives also a finite one--loop contribution
\cite{BGKM91,EM94}.

As a final observation, we note that all terms with fixed coefficients
in (\ref{piN3}) except one contain the spin matrix $S$. The odd term
without an explicit spin matrix has at least one external gauge
field through the tensor field $f_-^{\mu\nu}$.

\paragraph{6.} By suitably chosen nucleon field transformations, we have
brought the pion--nucleon Lagrangian of $O(p^3)$ in HBCHPT to a standard form
\beq
\wh \cL_{\pi N}^{(1)} + \wh \cL_{\pi N}^{(2)} + \wh \cL_{\pi N}^{(3)}~,
\label{piNt}
\eeq
with the three Lagrangians given by Eqs. (\ref{piN1}), (\ref{piN2}) and
(\ref{piN3}). In addition to $g_A$ and $m$, this Lagrangian  contains
7 scale--independent LECs $a_i$ of $O(p^2)$ and 24 renormalized, in
general scale--dependent LECs $b_i^r(\mu)$ of $O(p^3)$. By construction,
it is fully Lorentz invariant despite the dependence on an arbitrary
four--vector $v$. All equation--of--motion terms have been transformed away.
Consequently, all 1PI vertices can be read off directly from (\ref{piNt}).
More work is needed to determine and to interpret the coupling constants of
the $O(p^3)$ Lagrangian.

\vfill

\section*{Acknowledgements}
\noindent We are grateful to Gilberto Colangelo and J\"urg
Gasser for helpful discussions and to Veronique Bernard and Ulf Mei\ss ner
for a useful correspondence. G.E. thanks Pedro Pascual and the Benasque
Center of Physics (Spain) for the kind hospitality during the
final stage of this work. M.M. thanks the members of the
Institut f\"ur Theoretische Physik, Universit\"at Wien for their
warm hospitality.

\newpage

\newcommand{\PL}[3]{{Phys. Lett.}        {#1} {(19#2)} {#3}}
\newcommand{\PRL}[3]{{Phys. Rev. Lett.} {#1} {(19#2)} {#3}}
\newcommand{\PR}[3]{{Phys. Rev.}        {#1} {(19#2)} {#3}}
\newcommand{\NP}[3]{{Nucl. Phys.}        {#1} {(19#2)} {#3}}

\end{document}